\documentclass{ifacconf}

\usepackage{comment}
\usepackage{graphicx}      
\usepackage{natbib}        
\usepackage{amsmath,amssymb,amsfonts}     
\begin{document}
\begin{frontmatter}

\title{Backstepping Control of PDEs on Domains with
Graph-Monotone Boundaries}

\thanks[footnoteinfo]{This extended abstract has been submitted to the 2026 International Symposium on Mathematical Theory of Networks and Systems (MTNS 2026), to be held in Waterloo, ON, Canada.}

\author{Mohamed Camil Belhadjoudja} 

\address{Department of Applied Mathematics, University of Waterloo, 200 University Avenue West, Waterloo, ON, Canada, N2L 3G1\\
(e-mail: m2camilb@uwaterloo.ca)}

\end{frontmatter}
Phenomena that can be modeled by one-dimensional partial differential equations (PDE)s are relatively rare. Such models typically arise only when the process evolves in domains with special geometries and when certain dynamics are neglected. Examples include the temperature of fluids flowing through concentric cylinders in heat exchangers and the density of vehicles along a roadway. Outside these specific configurations, and even for problems as elementary as heat diffusion in a thin plate, higher-dimensional PDEs must be considered. The present work focuses on boundary control of such higher-dimensional systems via the backstepping approach.

Despite the extensive body of work on backstepping for one-dimensional PDEs, as surveyed in \cite{vaz_1}, results in higher dimensions remain comparatively limited. Most available methods either exploit particular symmetries of the PDE \cite{vaz_2,vaz_3,vaz_4,vaz_5,vaz_6,vaz_7,vaz_8,liu} or address problems posed on parallelepiped domains \cite{meurer}. To the best of our knowledge, the only approach that enables the design of backstepping controllers on non-parallelepiped regions without symmetry assumptions is the domain extension technique introduced in \cite{vaz_key1,vaz_key2}. This method, however, presents several drawbacks. In particular, the control input at each time instant is obtained by simulating a PDE on an extended domain, from which the actual input on the original domain is approximated. By contrast, in the one-dimensional setting, once the time-independent backstepping gain kernel is known, the control input can be computed in closed form as a feedback depending solely on the state at that same instant. Moreover, problems such as output-feedback design or adaptive and robust control do not appear straightforward to address with the domain extension method, at least to the best of our knowledge. These considerations motivate the search, whenever possible, for alternatives that preserve the main advantages of one-dimensional backstepping.

A motivating example for the domain extension method in \cite{vaz_key1,vaz_key2} is the control of the heat equation on a piano-shaped domain, with actuation applied at the tail of the piano. In this extended abstract, we show through a simple calculation that the domain extension method is not required in this setting. Instead, a strategy akin to that used for parallelepiped domains can be adopted. This result constitutes a first instance of a broader framework for backstepping control of asymmetric PDEs posed on non-parallelepiped regions, which we refer to as domains with \textit{graph-monotone boundaries}. The general framework is developed in \cite{camil}.

\begin{figure}
    \centering
    \includegraphics[width=\linewidth]{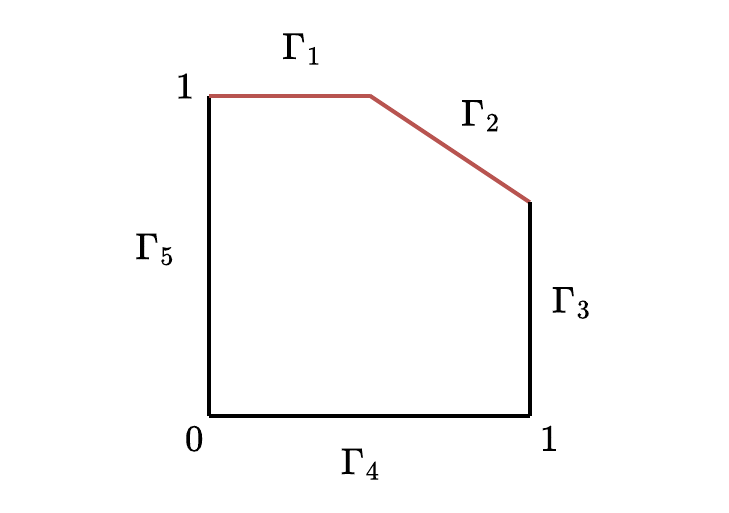}
    \caption{The piano-shaped domain $\Omega$. The tail of the piano is the boundary $\Gamma_1\cup \Gamma_2$. The $x$-coordinate is the horizontal one, and the $y$-coordinate is the vertical one.}
    \label{fig1}
\end{figure}

Consider the heat equation
\begin{align}
v_t = v_{xx}+v_{yy}+\lambda v, \label{one}
\end{align}
where $v(x,y,t)\in \mathbb{R}$ is the state variable, $t\geq 0$ is the time variable, $(x,y)\in \Omega$ is the space variable with $\Omega$ the piano-shaped domain depicted in Figure \ref{fig1}, and $\lambda\in \mathbb{R}$ is the reaction coefficient.

The following Dirichlet-type boundary conditions are imposed 
\begin{align*}
&v(x,y,t) = 0 \quad &&\text{on $\left(\bigcup_{i=3}^{5}\Gamma_i\right)\times [0,+\infty)$}, \\
&v(x,1,t) = U_1(x,t)\quad &&\text{on $\Gamma_1\times [0,+\infty)$}, \\
&v(x,y,t) = U_2(x,y,t)\quad &&\text{on $\Gamma_2\times [0,+\infty)$},
\end{align*}
where $U_1(x,t),U_2(x,y,t)\in \mathbb{R}$ are control variables to be designed in order to stabilize the origin $\{v=0\}$ in the sense of the $L^2$ norm. 

To solve the above control-design problem without using the domain extension method, we propose to consider the backstepping transformation 
\begin{align*}
w(x,y,t) := v(x,y,t) - \int_{0}^{y}K(y,\xi)v(x,\xi,t)d\xi,
\end{align*}
for all $(x,y)\in \Omega$ and $t\geq 0$. Note that, for each $x\in [0,1]$, the kernel $K$ must be defined on a domain that includes
\begin{align*}
\mathcal{T}(x) := \{(y,\xi)\in \mathbb{R}^2:0\leq \xi\leq y\leq \phi(x)\},
\end{align*}
where $\{(x,\phi(x)) \in \mathbb{R}^2:x\in [0,1]\}$ is the graph of the control boundary $\Gamma_1\cup \Gamma_2$. Since $\mathcal{T}(x)\subset \mathcal{T}$, where 
\begin{align*}
\mathcal{T} := \{(y,\xi)\in \mathbb{R}^2:0\leq \xi\leq y\leq 1\},
\end{align*}
then we can choose $K$ to be defined on $\mathcal{T}$, so that the domain of the kernel is independent of the $x$ variable. Then, the objective is to select $K$, $U_1$ and $U_2$ so that $w$ solves the exponentially-stable target system
\begin{align}
&w_t = w_{xx}+w_{yy} \quad &&\text{on int$(\Omega)\times (0,+\infty)$},\label{1_w} \\
&w = 0 \quad &&\text{on $\partial \Omega \times [0,+\infty)$},\label{2_w}
\end{align}
where int$(\Omega)$ (resp., $\partial \Omega$) is the interior (resp., the boundary) of $\Omega$.

Differentiating $w$ with respect to time and using \eqref{one}, we obtain 
\begin{align*}
w_t(x,y) =&~ v_t(x,y)-\int_{0}^{y}K(y,\xi)v_t(x,\xi)d\xi\nonumber \\
=&~ v_{xx}(x,y)+v_{yy}(x,y)+\lambda v(x,y)\nonumber \\
&~ - \lambda \int_{0}^{y}K(y,\xi)v(x,\xi)d\xi \nonumber \\
&~ - \int_{0}^{y}K(y,\xi)v_{xx}(x,\xi)d\xi \nonumber \\
&~-\int_{0}^{y}K(y,\xi)v_{\xi\xi}(x,\xi)d\xi. 
\end{align*}
Using integration by parts, note that we have 
\begin{align*}
\int_{0}^{y}K(y,\xi)v_{\xi\xi}(x,\xi)&d\xi = K(y,y)v_y(x,y)-K(y,0)v_y(x,0) \nonumber \\
&~\qquad - \int_{0}^{y}K_{\xi}(y,\xi)v_\xi(x,\xi)d\xi \nonumber \\
=&~K(y,y)v_y(x,y)-K(y,0)v_y(x,0) \nonumber \\
&~-K_{\xi}(y,y)v(x,y)+K_{\xi}(y,0)v(x,0) \nonumber \\
&~+\int_{0}^{y}K_{\xi\xi}(y,\xi)v(x,\xi)d\xi. 
\end{align*}
Next, differentiating $w$ twice with respect to $x$, we obtain 
\begin{align*}
w_{xx}(x,y) = v_{xx}(x,y)-\int_{0}^{y}K(y,\xi)v_{xx}(x,\xi)d\xi.
\end{align*}
Similarly, differentiating $w$ twice with respect to $y$, we have 
\begin{align*}
w_{yy}(x,y) =&~ v_{yy}(x,y)-\frac{\partial}{\partial y}\bigg(K(y,y)v(x,y) \nonumber \\
&~+\int_{0}^{y}K_y(y,\xi)v(x,\xi)d\xi\bigg) \nonumber \\
=&~ v_{yy}(x,y) - \left(\frac{d}{dy}K(y,y)\right)v(x,y)\nonumber \\
&~-K(y,y)v_y(x,y)-K_y(y,y)v(x,y) \nonumber \\
&~-\int_{0}^{y}K_{yy}(y,\xi)v(x,\xi)d\xi.
\end{align*}
As a result, we obtain 
\begin{align*}
w_t(x,&y) = w_{xx}(x,y)+w_{yy}(x,y)+K(y,0)v_y(x,0) \nonumber \\
&~+\left(\lambda + 2\frac{d}{dy}K(y,y)\right)v(x,y)\nonumber \\
&~+\int_{0}^{y}v(x,\xi) \bigg(K_{yy}(y,\xi)-K_{\xi\xi}(y,\xi)-\lambda K(y,\xi)\bigg)d\xi.
\end{align*}
Hence, to enforce \eqref{1_w}, it is sufficient to let $K$ solve the kernel equations 
\begin{align*}
&K_{yy}-K_{\xi\xi}-\lambda K = 0, \\
&K(y,y) = -\frac{\lambda}{2}y, \\
&K(y,0) = 0,
\end{align*}
on int$(\mathcal{T})$. These are exactly the same kernel equations as in the one-dimensional setting \cite{first_p}. In particular, since $\lambda$ is constant, then $K$ can be obtained explicitly as 
\begin{align*}
K(y,\xi) = -\lambda \xi \frac{I_1\left(\sqrt{\lambda (y^2-\xi^2)}\right)}{\sqrt{\lambda (y^2-\xi^2)}},
\end{align*}
where $I_1$ is the modified Bessel function of order $1$. 

Finally, to enforce the boundary condition \eqref{2_w}, it is sufficient to let 
\begin{align*}
U_1(x,t) =&~ \int_{0}^1K(1,\xi)v(x,\xi,t)d\xi \quad \text{on $\Gamma_1\times [0,+\infty)$}, \\ 
U_2(x,y,t) =&~ \int_{0}^{\phi(x)}K(\phi(x),\xi)v(x,\xi,t)d\xi \\
&~\qquad \qquad \qquad \qquad \qquad \quad \text{on $\Gamma_2\times [0,+\infty)$.} 
\end{align*}

In \cite{camil}, we identify the key properties of the heat equation on the piano-shaped domain that make the above calculations possible. We then abstract these features to a class of domains in $\mathbb{R}^n$, $n>1$, with so-called graph-monotone boundaries, and develop backstepping controllers on such domains in a way that parallels the one-dimensional setting. In particular, we show that results previously obtained for parallelepiped geometries can be extended to a variety of "amorphous" regions by enlarging the domain of the backstepping kernel rather than that of the PDE itself, while restricting the integration region appearing in the control law.


\begin{thebibliography}{xx}  

\bibitem[Belhadjoudja(2026)]{camil}
M. C. Belhadjoudja, 
\newblock Backstepping Control of PDEs on Domains with Graph-Monotone Boundaries
\newblock \emph{Paper in prepartion.}

\bibitem[Liu and Xie(2020)]{liu}
X. Liu and C. Xie,
\newblock Boundary control of reaction–diffusion equations on higher-dimensional symmetric domains, \newblock \emph{Automatica}, 114, 108832, 2020.

\bibitem[Meurer(2016)]{meurer}
T. Meurer,
\newblock Control of Higher-Dimensional PDEs: Flatness and Backstepping Designs,
\newblock \emph{Springer}, 2013.

\bibitem[Smyshlyaev and Krstic(2004)]{first_p}
A. Smyshlyaev and M. Krstic,
\newblock Closed-form boundary state feedbacks for a class of 1-D partial integro-differential equations,
\newblock \emph{IEEE Transactions on Automatic control}, 49(12), 2185-2202, 2004.

\bibitem[Vazquez(2024)]{vaz_key1}
R. Vazquez, 
\newblock Exploiting Symmetry in Higher-Dimensional PDE Control: A Backstepping Perspective,
\newblock \emph{Control and Adaptation: Imagine What’s Next - Celebrating the 60th Birthday of Miroslav Krstic}, 2024.

\bibitem[Vazquez(2025)]{vaz_key2}
R. Vazquez,
\newblock Backstepping control laws for higher-dimensional PDEs: spatial invariance and domain extension methods,
\newblock \emph{IMA Journal of Mathematical Control and Information}, 42(2), dnaf018, 2025.

\bibitem[Vazquez et al.(2026)]{vaz_1}
R. Vazquez, J. Auriol, F. Bribiesca-Argomedo, M. Krstic,
\newblock Backstepping for partial differential equations: A survey,
\newblock \emph{Automatica}, Volume 183, 112572, 2026.

\bibitem[Vazquez and Krstic(2007)]{vaz_2}
R. Vazquez and M. Krstic,
\newblock A closed-form feedback controller for stabilization of the linearized 2-D Navier–
Stokes Poiseuille system,
\newblock \emph{IEEE Transactions on Automatic Control}, 52(12):2298–2312, 2007.

\bibitem[Vazquez and Krstic(2010)]{vaz_3} 
R. Vazquez and M. Krstic,
\newblock Boundary observer for output-feedback stabilization of thermal convection loop,
\newblock \emph{IEEE Transactions on Control Systems Technology}, 18:789–797, 2010.

\bibitem[Vazquez and Krstic(2016)]{vaz_4}
R. Vazquez and M. Krstic,
\newblock Boundary control of a singular reaction-diffusion equation on a disk,
\newblock \emph{IFAC-
PapersOnLine}, 49(8):74–79, 2016.

\bibitem[Vazquez and Krstic(2019)]{vaz_5}
R. Vazquez and M. Krstic,
\newblock Boundary control and estimation of reaction-diffusion equations on the sphere under revolution symmetry conditions,
\newblock \emph{International Journal of Control}, 92(1):2–11, 2019.

\bibitem[Vazquez et al.(2008)]{vaz_6}
R. Vazquez, E. Schuster, and M. Krstic, \newblock Magnetohydrodynamic state estimation with boundary sensors,
\newblock \emph{Automatica}, 44(10):2517–2527, 2008.

\bibitem[Vazquez et al.(2023)]{vaz_7}
R. Vazquez, J. Zhang, J. Qi, and M. Krstic,
\newblock Kernel well-posedness and computation by power series in backstepping output feedback for radially-dependent reaction-diffusion PDEs on multidimensional balls,
\newblock \emph{Systems \& Control Letters}, 177:105538, 2023.

\bibitem[Xu et al.(2008)]{vaz_8}
C. Xu, E. Schuster, R. Vazquez, and M. Krstic, \newblock Stabilization of linearized 2D magnetohydrodynamic channel flow by backstepping boundary control,
\newblock \emph{Systems \& Control Letters}, 57(10):805–812, 2008.







\end{thebibliography}
\end{document}